\documentclass[conference]{IEEEtran}
\ifCLASSINFOpdf
\else
\fi
\usepackage{cite}
\usepackage{graphicx}
\usepackage[small]{caption}
\usepackage{subcaption}
\usepackage{color,soul}
\usepackage{array}
\usepackage{amsmath}
\usepackage{booktabs}
\usepackage{footnote}
\makesavenoteenv{tabular}
\makesavenoteenv{table}
\usepackage{color}
\usepackage{amssymb}
\usepackage{algorithm}
\usepackage[noend]{algpseudocode}

\begin{document}
\pagestyle{empty}
\title{SFQmap: A Technology Mapping Tool for Single Flux Quantum Logic Circuits}

\author{Ghasem Pasandi, Alireza Shafaei and Massoud Pedram\\
  Department of Electrical Engineering-Systems\\ University of Southern California, Los Angeles, CA 90089.

}

%

\maketitle

\begin{abstract}
Single flux quantum (SFQ) logic is a promising candidate to replace the CMOS logic for high speed and low power applications due to its superiority in providing high performance and energy efficient circuits. However, developing effective Electronic Design Automation (EDA) tools, which cater to special characteristics and requirements of SFQ circuits such as depth minimization and path balancing, are essential to automate the whole process of designing large SFQ circuits. In this paper, a novel technology mapping tool, called SFQmap, is presented, which provides optimization methods for minimizing first the circuit depth and path balancing overhead and then the worst-case stage delay of mapped SFQ circuits. Compared with the state-of-the-art technology mappers, SFQmap reduces the depth and path balancing overhead by an average of 14\% and 31\%, respectively.
\end{abstract}

\IEEEpeerreviewmaketitle

\section{Introduction}
\label{Intro:sec}
Logic synthesis is an important step in the design flow of digital circuits and systems with a big impact on the final delay and area of the chip. Logic synthesis has two main phases: \textit{technology-independent} optimizations, and \textit{technology mapping}. The first phase includes several algebraic or Boolean optimizations such as restructuring, resubstitution, common subexpression extraction, node minimization. The second phase performs  mapping or binding of logic expressions to actual gates in a given cell library to minimize the circuit delay, area or other desired metrics. 

Rapid SFQ (RSFQ) is a family of  Single Flux Quantum circuits, which was developed in late 1980s \cite{likharev1991rsfq}. More recent versions of SFQ logic family include ERSFQ \cite{kirichenko12ersfq,mukhanov1987ultimate} and eSFQ \cite{mukhanov2011energy} as well as AQFP \cite{takeuchi2014energy}. Generally, SFQ logic has been touted as a good candidate for achieving energy efficient and high performance circuits \cite{holmes2013energy}. SFQ devices are made of Josephson Junctions, which are superconducting devices that exhibit the Josephson effect, i.e., the  phenomenon of a current (called super-current) that flows indefinitely long without any applied voltage. The propagation delay through these devices is as low as $1ps$, while each switching action consumes in the order of $10^{-19}J$ energy  \cite{bunyk1995high}. The JJ switching energy is 2-3 orders of magnitude lower than that of the end-of-CMOS-scaling devices as stated in \cite{Theis_Wong_MooresLaw}, which again shows the promise of SFQ circuits as a replacement for CMOS circuits. SFQ gates can operate as fast as 370GHz at $T=4.2K$ \cite{bunyk1995high}. 

Although advantages of SFQ logic in achieving super fast and very low-power circuits have been proven, state of the art in Electronic Design Automation (EDA) technologies is not advanced.  On the other hand, due to  key differences between SFQ and CMOS logic, CMOS Computer Aided Design (CAD) tools cannot be directly used to optimize SFQ circuits. Therefore, it is critical to develop appropriate CAD tools including synthesis, place and route, timing and power analysis, and verification tools to fully automate the design and test process of SFQ circuits.

In this paper, we present SFQmap, a technology mapping tool for SFQ circuits. To the best of authors' knowledge, this is the first paper which presents a technology mapping tool designed specifically for optimizing SFQ circuits. There are some tools for logic synthesis of SFQ, but they either provide simple optimizations or are based on standard CMOS CAD tools with minor pre or post optimizations (Section \ref{Prior-Work}). SFQmap includes two main phases: (i) logical depth minimization with path balancing, and (ii) peephole optimization to reduce the worst-case delay of any stage of the logic circuit. SFQmap improves key parameters of SFQ circuits considerably compared with the state-of-the-art technology mappers. For example, it decreases the number of path balancing D-Flip-Flops (DFFs), and the Product of the worst-case Stage delay with the logical Depth of the circuit (which we shall denote as PSD) by up to 43\%, and 63\%, respectively (Section \ref{exper:sec}). 

\section{Background}
\label{Prior-Motiv:sec}
\subsection{Background on SFQ (SFQ Basics)}
\label{BG-SFQ:sec}
Single Flux Quantum (SFQ) logic uses a single quantum of magnetic flux ($\Phi_0 = h/2e = 2.07 mV\times ps$) to represent the binary information. This SFQ typically appears as a voltage pulse with a peak amplitude of 2-4mV and a duration of 1-2ps. In this representation, the presence of a SFQ pulse is considered as ``logic-1'', while the absence of the pulse is interpreted as ``logic-0'' \cite{gross2015applied}. In the following, we explain characteristics and key circuit/gate level requirements of SFQ logic.

\subsubsection{Fanout in SFQ}\label{FanoutSubSubSec}
\begin{figure}[t]
        \centering
        \begin{subfigure}[!t]{0.23\textwidth}
                \centering
                \includegraphics[width=\textwidth]{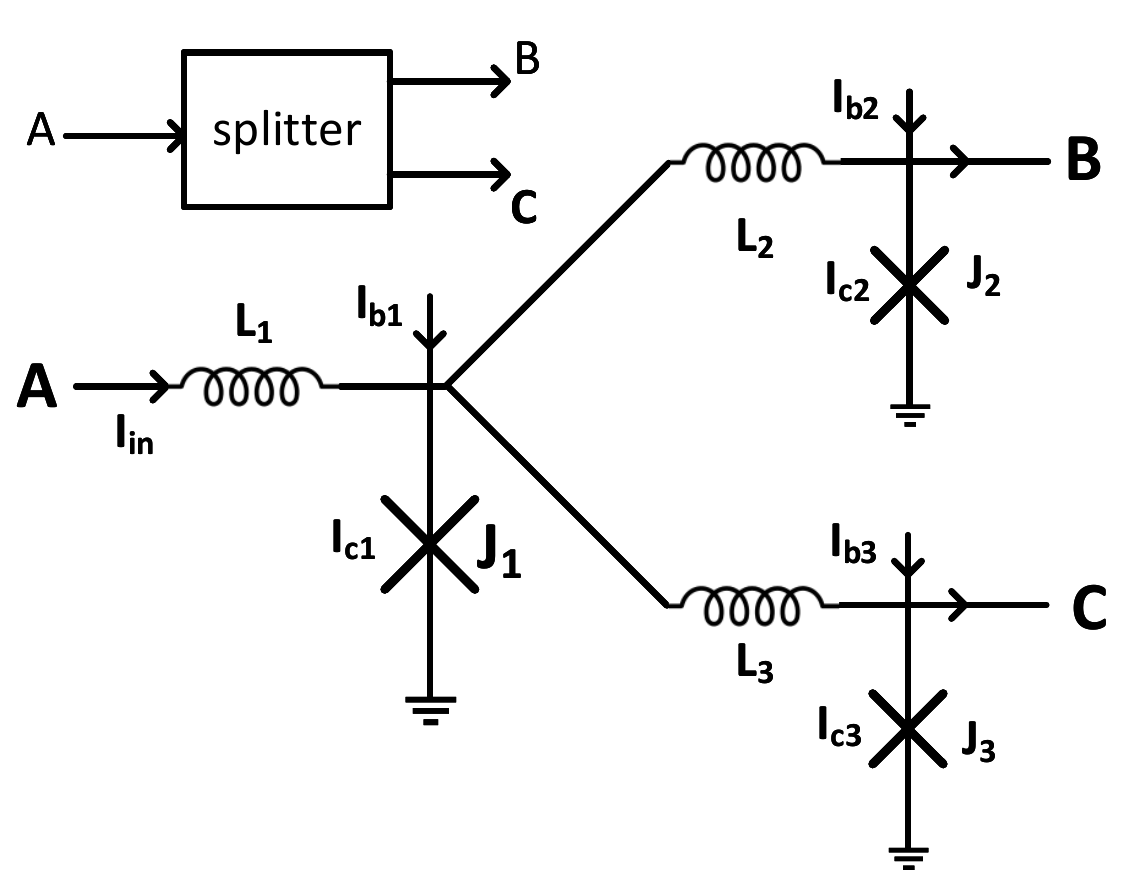}
                \caption{}
                \label{6T_cell}
        \end{subfigure}
        \begin{subfigure}[!t]{0.23\textwidth}
                \centering
                \includegraphics[width=\textwidth]{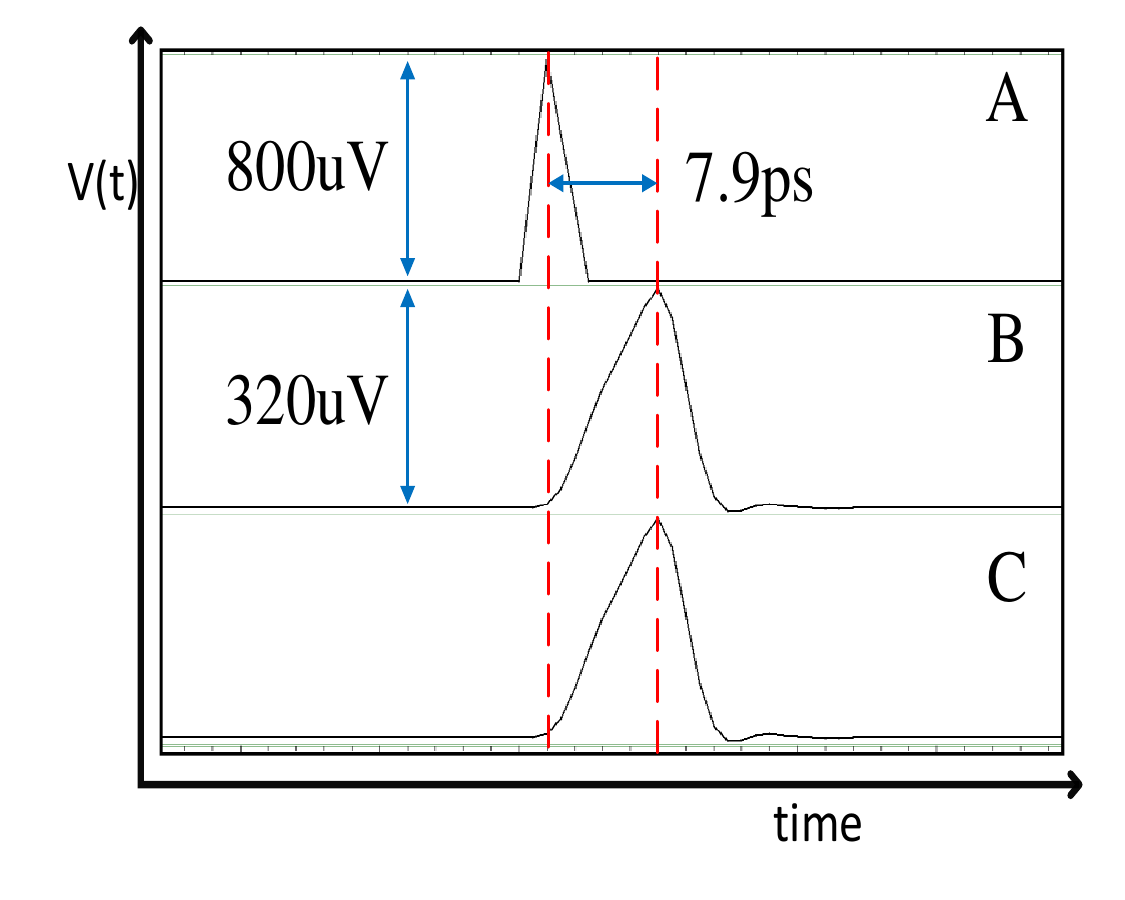}
                \caption{}
                \label{6T_Layout}
        \end{subfigure}
        \begin{subfigure}[!t]{0.3\textwidth}
                \centering
                \includegraphics[width=\textwidth]{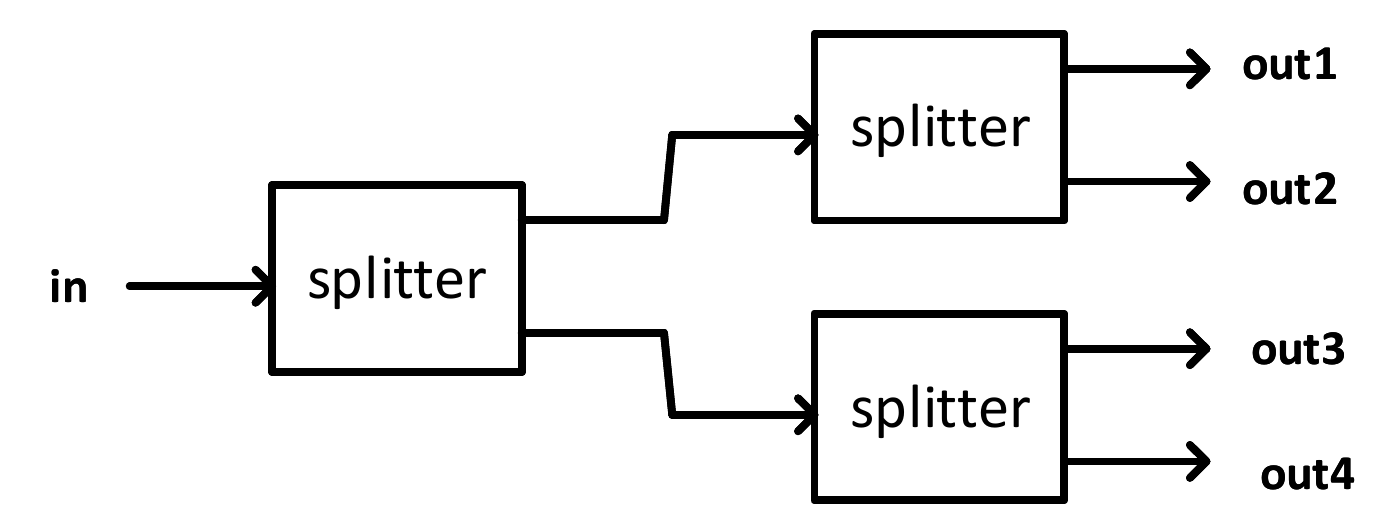}
                \caption{}
                \label{6T_Layout}
        \end{subfigure}
        \caption{(a) Splitter gate in SFQ electronics, (b) the waveform corresponding to the operation of the splitter gate, and (c) a splitter tree to provide FO4. The waveform is borrowed from \cite{katam2017design}.}
        \label{Splitter}
\end{figure}

In conventional SFQ logic, each gate can only drive one other node (fanout count is one). If a gate needs to have more than one fanout, an special SFQ gate called a splitter cell is inserted at the output of that gate to enable fanout to two gates. To support additional fanouts, a binary tree of splitter cells can be used. Fig. \ref{Splitter} shows the circuit-level schematic of a splitter gate, its corresponding waveform, and an example of a splitter tree for fanout of 4 (FO4).
\subsubsection{Gate-level pipeline}\label{pipeSubSubSec}All SFQ gates, except splitters, require a clock signal to transfer the stored quantum flux to their output. In addition,  operation of all SFQ gates is synchronized by the clock. For distribution of the clock signal in a SFQ chip, there are three main methods: \textit{counter-flow clocking} in which the clock flows in the opposite direction of the data,  \textit{concurrent-flow clocking}, where clock and data flow in the same direction, and \textit{clock-follow-data}, in which the clock arrives after the data input to a gate have arrived and been processed by the gate. For more information on the clock distribution network of the SFQ circuits, please refer to \cite{friedman2001clock,katam2017design}. The take-away  is that nearly all SFQ gates (with the exception of some asynchronous gates such as a splitter or a Josephson Transmission Line) can be thought of as  purely combinational gates followed by  clocked DFFs. The combinational gate responds to its inputs by changing its internal loop current state whereas the gate output  will  change only after the clock pulse comes, which will in turn reset the gate's internal state  while producing the correct gate output value. In other words, every SFQ circuit must be completely gate-level pipelined. 

\subsubsection{Path Balancing}\label{PathBalancingSubSubSec} Due to gate-level pipelining in SFQ, for correct operation of the SFQ gates, all inputs of a gate should have the same logic level\footnote{Logic level of gate $i$ denotes the length of the longest path (in terms of the gate count) from any Primary Input (PI) of the network to this gate.}. If inputs of a gate do not have the same logic level, some DFFs should be inserted into the path with smaller depth to balance the network. For example, if the first input ($in_1$) in an AND2 gate has the logic level of 2 and the second input ($in_2$) has the logic level of 3, one DFF should be added to $in_1$. This path balancing is needed because once the clock arrives at a gate, the available input data (which is stored as a loop current in the gate) is reset in order to produce the correct output value. So, if inputs to a gate arrive too early or too late, the gate will produce wrong results when  corresponding clock pulses arrive at the gate. In the above example, if $in_1$ and $in_2$ are both logic 1 values,  $in_1$ arrives early, but $in_2$ does not arrive until after the clock pulse has come in (the AND2 gate will produce a wrong output result of logic 0 in response to the clock pulse.)

\subsubsection{Product of Stage Delay and Circuit Depth}\label{ClockSubSubSec} Due to the gate-level pipelined nature of the SFQ logic, the input-to-output latency of a SFQ circuit is determined as the product of the clock cycle time and the logical depth of the circuit. The clock cycle time is in turn set by the worst-case delay of any single stage (gate plus any present splitters plus interconnect) of logic in the circuit. 
\section{Technology Mapping}
\label{DP_Tech-Map}
\subsection{Prior Work}
\label{Prior-Work}
In \cite{keutzer1987dagon,chaudhary1995computing}, a tree mapping and decomposition method is presented to generate minimum area or low power circuits.  In \cite{chen2004daomap}, a cut-enumeration-based method which involves cut generation and cut selection was presented targeting the depth-optimal area optimization for the FPGAs.
For the SFQ circuits, there are couple of papers addressing the logic synthesis \cite{yamashita2006transduction,yoshikawa2001top,katam2017desig_isec}. In \cite{yamashita2006transduction}, a framework is proposed for synthesizing the SFQ circuits by constructing a virtual cell called ``2-AND/XOR". In \cite{yoshikawa2001top}, a top-down design methodology for the SFQ circuits based on the Binary Decision Diagram (BDD) is presented. In \cite{katam2017desig_isec}, an academic logic synthesis tool (ABC)\cite{synthesis2011abc} is modified to meet the requirements of SFQ logic without developing any SFQ specific optimization algorithms for the technology mapping.
\subsection{Motivation}
\label{Motiv:sec}
\begin{figure}[t]
\centering
\includegraphics[width=0.5\textwidth]{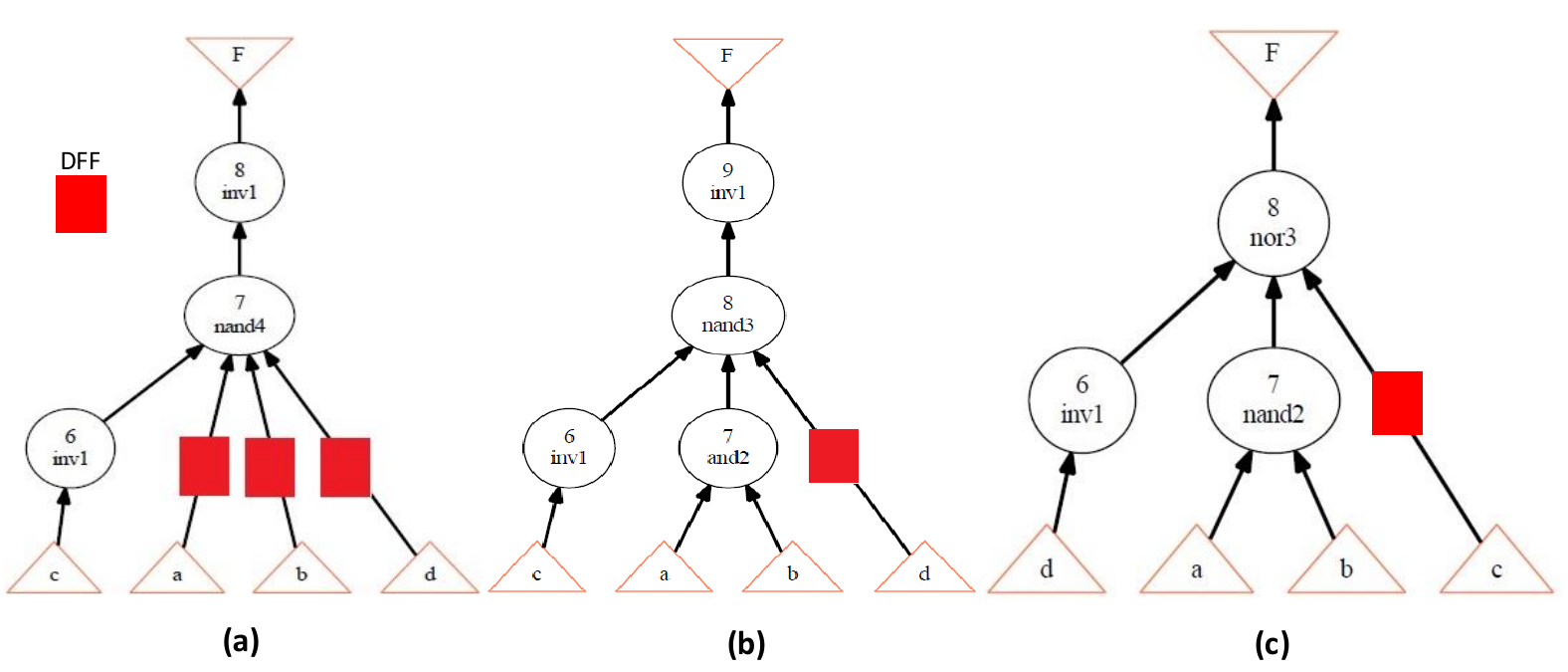}
\caption{Three mapping solutions for the expression $F = a \cdot b \cdot (!c) \cdot d$. (a) the circuit generated by ABC mapper \cite{synthesis2011abc} requiring three path balancing DFFs and has depth of three; two other mapping solutions requiring only one DFF with depth of (b) three, and (c) two.}
\label{Motive_example}
\end{figure}

To map the Boolean expression, $F = a \cdot b \cdot (!c) \cdot d$, the ABC mapper \cite{synthesis2011abc} produces the circuit shown in Fig. \ref{Motive_example}(a). As explained in Section \ref{PathBalancingSubSubSec}, the SFQ circuits need to be path balanced. Thus, for the mapped circuit generated by ABC, three path balancing DFFs have to be inserted into the network. Another circuit with fewer number of required path balancing DFFs can be found to implement the given Boolean expression, as shown in Fig. \ref{Motive_example}(b). On the other hand, as discussed in Section \ref{ClockSubSubSec}, to reduce the computation latency, the PSD should be minimized. In practice, between the logical depth and the worst-case stage delay, the former has more impact on the overall computation latency; therefore, we target the minimization of the logical depth (along with minimization of the path-balancing DFF overhead) first; Only after this optimization  is done we turn to direct minimization of the product of the stage delay and logical depth.  Based on this explanation, the circuit shown in Fig. \ref{Motive_example}(c) is more desirable than the other two . This is because it requires only one DFF while its logical depth is less. We need to develop a technology mapper for SFQ which prefers the circuit shown in Fig. \ref{Motive_example}(c) over the other two mapping solutions.
\subsection{Algorithms}
\label{our-rsfq-sub-sec}
Our proposed technology mapping for the SFQ circuits consists of two main phases: \textit{depth minimization and path balancing}, and \textit{peephole optimization for reducing the PSD}.
The aforesaid phases are discussed in more details in the following.
\subsubsection{Depth Minimization with Path Balancing}
\label{DM:subsec}
In this sub-section, we present the problem of tree-mapping as a Dynamic Programming (DP) problem. The optimal solution is the one with the least depth and in the case of a tie, it is the one which requires less number of path balancing DFFs. A similar cut-enumeration methods as in \cite{cong1994flowmap, mishchenko2007combinational} is used. As mentioned in \cite{cong1994flowmap}, this method provides the optimal depth solution for general DAGs. The main difference between the method which is presented in \cite{cong1994flowmap} and our depth minimization and path balancing method is that we optimize both depth and balancing, while in \cite{cong1994flowmap} only depth is considered.

\begin{figure}[t]
\centering
\includegraphics[width=0.33\textwidth]{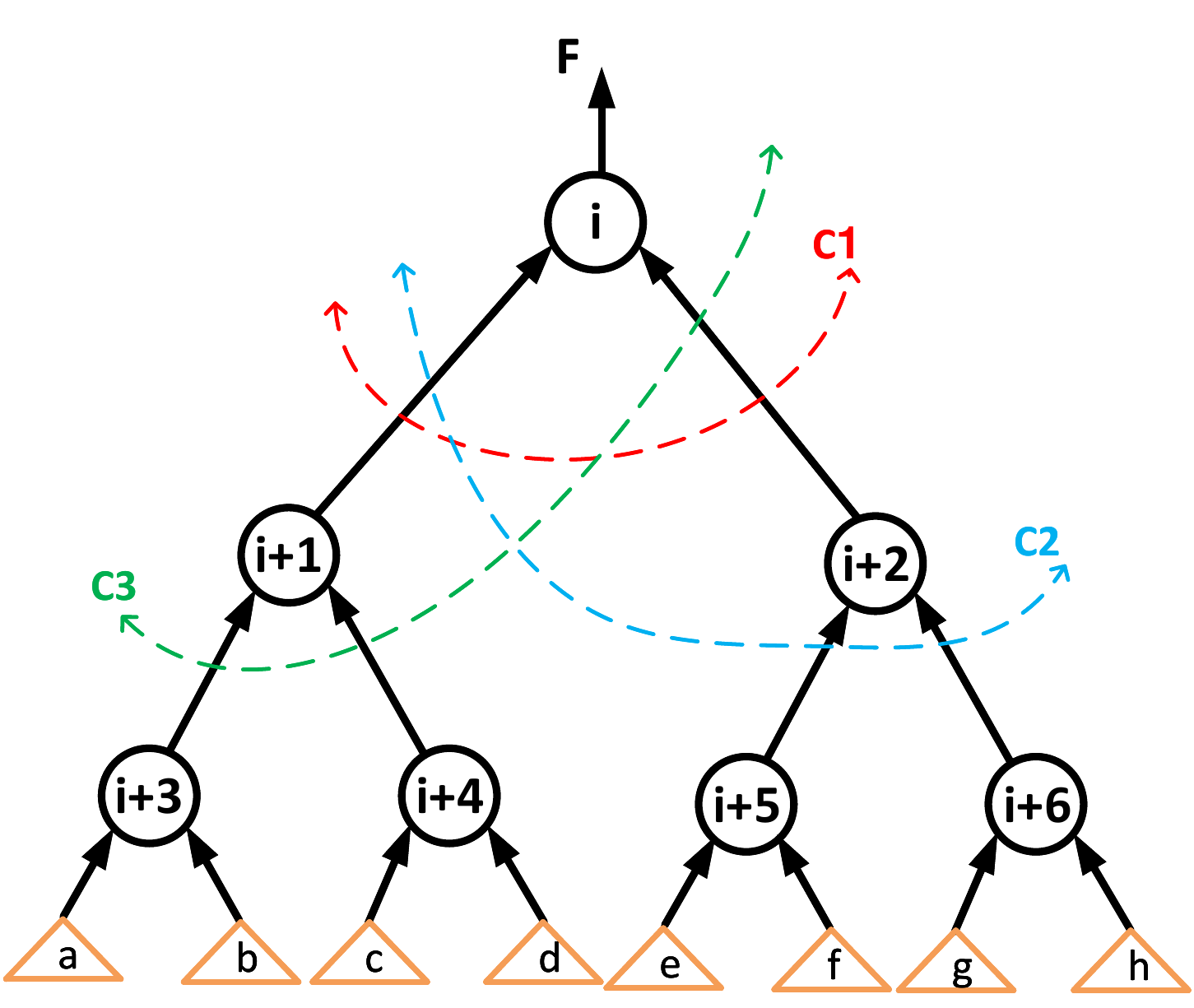}
\caption{Showing $3$-$feasible$ cuts of node i: C1 to C3.}
\label{3-feasible-cuts}
\end{figure}

Fig. \ref{3-feasible-cuts} shows a binary tree and the $3$-$feasible$ cuts for node $i$. (refer to \cite{cong1994flowmap} for the definition of $k$-$feasible$ cuts.) Eq. \ref{OPT_depth_eq} defines the value of optimal depth solution, $D[i]$, as the minimum achievable logical depth for mapping a tree rooted at node $i$. $D[i]$ is calculated recursively as follows:
{\small
\begin{align}
\label{OPT_depth_eq}
D[i] & = min \lbrace \nonumber \\
& \quad  max\textbf{(} \quad D[i+1] , D[i+2] \quad \textbf{)} \quad + \quad 1\textbf{,} \nonumber \\
& \quad  max\textbf{(}\quad D[i+1] , D[i+5] , D[i+6] \quad \textbf{)} \quad + \quad 1\textbf{,} \nonumber \\
& \quad  max\textbf{(} \quad D[i+2] , D[i+3] , D[i+4] \quad \textbf{)} \quad + \quad 1\textbf{,} \nonumber \\
& \cdots \rbrace
\end{align}
}

The three terms in the Eq. \ref{OPT_depth_eq} corresponds to the depth of cuts $C1-C3$ (Fig. \ref{3-feasible-cuts}), respectively. Basically, in Eq. \ref{OPT_depth_eq} a mapping solution with the least depth ($D_{min}$) among all choices corresponding to the  $k$-$feasible$ cuts is being computed. Among all the solutions with depth $D_{min}$, the one which requires fewer path balancing DFFs is selected recursively in a Dynamic Programming approach as shown in Eq. \ref{OPT_DFF_eq}. In this equation, $DFF[i]$ gives the DFF count for a solution with depth of $D_{min}$ and the least number of path balancing DFFs (\#DFFs).\\
{\small
\begin{align}
& DFF[i] = min \lbrace \nonumber\\
& DFF[i+1] + DFF[i+2] + B(i+1,i+2)\textbf{,} \nonumber\\
& DFF[i+1] + DFF[i+5] + DFF[i+6] + B(i+1,i+5,i+6)\textbf{,} \nonumber\\ 
& DFF[i+2] + DFF[i+3] + DFF[i+4] + B(i+2,i+3,i+4)\textbf{,}\nonumber\\
& \cdots \rbrace
\label{OPT_DFF_eq}
\end{align}
}

In the above equation, $B()$ accounts for the required number of DFFs for balancing the inputs of the corresponding cut because of the level difference among cut's inputs. For example, if the level of node $i$+$1$ is $2$ and $i$+$2$ is $3$, $B(i$+$1,i$+$2)$ returns $1$.

The complexity of computing the $k$-$feasible$ cuts is $O(kmn)$, where $m$ is the edge count, and $n$ is the node count \cite{cong1994flowmap}. The complexity of our depth minimization and path balancing algorithm by having the $k$-$feasible$ cuts is $O(K'gn)$, where $K'$ is the maximum number of $k$-$feasible$ cuts for a node in the network, $g$ is the number of gates in the library, and $n$ is the total number of nodes in the network. Usually $K'$ and $g$ are small numbers. Thus, the complexity for this algorithm is determined by the number of nodes in the network with linear relationship ($= O(n)$).
\begin{algorithm}[t]
\caption{SFQmap}\label{SFQmap_alg}
\begin{algorithmic}[1]
{\scriptsize
\Procedure{Main-Procedure}{}
\State\textbf{input}: Network pNtk, \textbf{output}: mapped network pNtkMap
\State //perform pre-mapping computations:
\State Compute k-feasible cuts for pNtk;
\State Initialize the mapping manager, pMan to map the input network, pNtk;
\State // two phases of SFQmap:
\State Minimize\_Depth\_PBOverhead(pMan,pNtk);
\State pNtkNew = Tune\_PSD(pMan,pNtk); //peephole optimization
\State pNtkMap = NetworkFromMap (pMan, pNtkNew); //generating the mapped network
\State \textbf{return} pNtkMap;
\State
\hrulefill
\EndProcedure
\Procedure {Minimize\_Depth\_PBOverhead}{}
\State\textbf{input}: Network pNtk, \textbf{output}: depth-optimal and path balanced mapping solutions for nodes;
\For{\texttt{each node \textit{pNode} in network \textit{pNtk}}}
\State Find min-depth and balanced mapping solution based on Eqs. \ref{OPT_depth_eq},  \ref{OPT_DFF_eq}.
\EndFor
\State \textbf{end for}
\State
\hrulefill
\EndProcedure
\Procedure {Tune\_PSD}{}
\State\textbf{inputs}: Network pNtk, iterations $p$, \textbf{output}: Network pNtkNew with reduced PSD
\State Set(InitFanoutCount,MaxFanoutCount);
\State pNtkNew = Copy (pNtk)
\State FanoutCout = InitFanoutCount // initial fanout count 
\While {\texttt{p > 0}}
\If {\texttt{FanoutCount <= MaxFanoutCount}}
\State Find the node with worst stage delay, \textit{pNodeWorst};
\State Generate a network comprising pNodeWorst and its FanIOs, \textit{pNtkTemp};
\State Remap pNtkTemp subject to the FanoutCout limit;
\If {\texttt{PSD is reduced}}
\State Substitute pNtkTemp into pNtkNew;
\Else
\State Increase FanoutCount;
\EndIf
\State Decrease $p$ by 1;
\EndIf
\EndWhile
\State \textbf{end while}
\State \textbf{return} pNtkNew
\EndProcedure
}
\end{algorithmic}
\end{algorithm}
\begin{table*}[t]
  \centering
  \caption{Experimental results for SFQmap, and ABC mapper using mcnc.genlib library. The run-time, which measures the amount of time it takes to generate the mapping solution, is measured in second $s$.}
    \begin{tabular}{ccccccccccc}
    \toprule
     & \multicolumn{2}{c}{\#DFFs} & \multicolumn{2}{c}{\#Gates} &  \multicolumn{2}{c}{Logical Depth}  & \multicolumn{2}{c}{PSD} & \multicolumn{2}{c}{Run-time}\\
   Circuits   & SFQmap    & ABC   & SFQmap    & ABC      & SFQmap    & ABC  & SFQmap    & ABC & SFQmap    & ABC\\
    \toprule
s4863	&3381&	4274&	1183&	1010&		30&	36&	297&	378&	0.198&	0.15 \\
\midrule
c5315	&2519	&4437&	1399&	1206&		20&	25&	179&	234&	0.2&	0.12 \\
\midrule
c7552	&2603	&3639	&1786	&1457		&19	&21	&267.9	&718.2	&0.25	&0.22 \\
\midrule
s6669	&7621	&9638	&1517	&1462		&46	&52	&510.6	&842.4	&0.258	&0.145 \\
\midrule
s38417	&13953	&23253	&8363	&7471		&14	&16	&180.6	&252.8	&0.667&	0.32 \\
    \bottomrule
    \end{tabular}%
  \label{exp_table}%
\end{table*}%
\begin{table}[t]
  \centering
  \caption{Improvement percentages of ``SFQmap -i 5" and ``SFQmap -i 0" over ABC mapper for key parameters. These results are the average of all five tested benchmark circuits. $\downarrow$ shows a decrease, and $\uparrow$ shows an increase in the corresponding quantity.}
    \begin{tabular}{cccccc}
    \toprule
    Mapper & Logical Depth & \#DFFs  & PSD  & Run-time  \\
    \midrule
    SFQmap -i 5 & $\downarrow$ 14\% & $\downarrow$ 31\% & $\downarrow$ 35\%   & $\uparrow$ 34\%\\
    \midrule
    SFQmap -i 0 &  $\downarrow$ 15\% & $\downarrow$ 37\% & $\downarrow$ 13\%   & $\uparrow$ 12\%\\
    \bottomrule
    \end{tabular}%
  \label{exp_table_percent}%
\end{table}%
\subsubsection{Peephole Optimization for Reducing the Sequential Depth}
\label{PHO:subsec}
As explained in Section \ref{ClockSubSubSec}, despite in CMOS, in the SFQ circuits to reduce the computation latency, PSD has to be reduced. To reduce the PSD, we perform the following heuristic: 

After finding the minimum depth and most balanced solution for each node of the network as in Section \ref{DM:subsec}, a gate with the worst stage delay is found. This gate is usually a gate with high fanout count. Next, a temporary network consisting this gate and its \textit{immediate} fanins and fanouts is generated. The temporary network is re-mapped while the fanout count for any node is limited. If the product of the worst stage delay and the length of the longest path decreased, the move will be accepted. Otherwise, we increase the fanout counts' limit and re-do the process. This process is repeated $p$ times. Experimental results show that for having $p=5$, there will be a considerable decrease in the PSD of the circuit while the run-time is acceptable (Tables \ref{exp_table},\ref{exp_table_percent}).

The complexity of the peephole optimization is $O(m+n)$, where, $m$ is the edge count and $n$ is the node count of the network. This is because for calculating the fanout dependent stage delay for all nodes (to find the worst stage delay), a breath-first search should be done. The peephole optimization determines the overall complexity of the SFQmap tool to be $O(m+n)$.

Algorithm \ref{SFQmap_alg} describes the pseudo code of the SFQmap. In this algorithm, the main procedure as well as functions for combined depth minimization and path balancing, and then peephole optimization for reducing the PSD are shown.
\section{Experimental Results}
\label{exper:sec}
We implemented our SFQ specific technology mapping algorithms inside ABC \cite{synthesis2011abc}. We used several \textit{ISCAS} benchmark circuits \cite{hansen1999unveiling} for testing our developed technology mapper. The mcnc.genlib library is used. Table \ref{exp_table} lists the key parameters for different benchmark circuits for SFQmap and ABC technology mapping tools. As explained in Section \ref{our-rsfq-sub-sec}, our developed technology mapper focuses on improving three important parameters in SFQ circuits including \textit{logical depth}, \textit{\#DFFs}, and \textit{PSD}. As shown in Tables \ref{exp_table},\ref{exp_table_percent}, SFQmap provides considerable improvements on all of these critical parameters on average of all five benchmark circuits. However, its average run-time is increased by 34\% over the ABC mapper. This is mainly because of the peephole optimization phase which dominates the run-time of the SFQmap. We implemented the peephole optimization phase with the capability of determining the number of iterations. The experimental results in the  Table \ref{exp_table} is for five iterations of peephole optimization (\textit{SFQmap -i 5}). If we cross out the peephole optimization phase to trade the PSD with the run-time, the run-time overhead over the ABC mapper will be decreased to less than 12\%. In Table \ref{exp_table_percent}, ``\textit{SFQmap -i 0}" is for not having any peephole optimization runs. $i$ stands for the number of iterations. 
\section{Conclusion}
\label{conc:sec}
In this paper, a novel technology mapping tool, SFQmap, is presented which is developed for the SFQ circuits. This mapper performs two main optimizations to improve the most important parameters for the SFQ circuits including the logical depth, the number of path balancing DFFs, and the PSD. The implementation of this mapper allows to select the number of iterations for the optimization runs. ``SFQmap -i 5" improves the logical depth, \#DFFs, and the PSD by 14\%, 31\%, and 35\%, respectively over the ABC \cite{synthesis2011abc} mapper for five ISCAS benchmark circuits with 34\% increase in run-time. ``SFQmap -i 0" reduces the overhead of run-time to less than 12\% and provides even more improvements on the logical depth, and \#DFFs over ABC (see Table \ref{exp_table_percent}).

\section{Acknowledgement}
The research is based upon work supported by the Office of the Director of National Intelligence (ODNI), Intelligence Advanced Research Projects Activity (IARPA), via the U.S. Army Research Office grant W911NF-17-1-0120. The U.S. Government is authorized to reproduce and distribute reprints for Governmental purposes notwithstanding any copyright notation herein.

\ifCLASSOPTIONcaptionsoff
  \newpage
\fi
\bibliographystyle{IEEEtran}
\bibliography{IEEEabrv,template}

\end{document}